\documentclass[twocolumn,showpacs,preprintnumbers]{revtex4}%
\usepackage{amssymb}
\usepackage{amsmath}
\usepackage{graphicx}
\usepackage{dcolumn}
\usepackage{bm}
\usepackage{amsfonts}%
\begin{document}
\preprint{ }
\title{Quantum Gravity Boundary Terms from Spectral Action of Noncommutative Space}
\author{Ali H. Chamseddine$^{1,3}$and Alain Connes$^{2,3,4}$}
\affiliation{$^{1}$Physics Department, American University of Beirut, Lebanon}
\affiliation{$^{2}$College de France, 3 rue Ulm, F75005, Paris, France}
\affiliation{$^{3}$I.H.E.S. F-91440 Bures-sur-Yvette, France}
\affiliation{$^{4}$Department of Mathematics, Vanderbilt University, Nashville, TN 37240 USA}

\begin{abstract}
We study the boundary terms of the spectral action of the noncommutative
space, defined by the spectral triple dictated by the physical spectrum of the
standard model, unifying gravity with all other fundamental interactions. We
prove that the spectral action predicts uniquely the gravitational boundary
term required for consistency of quantum gravity with the correct sign and
coefficient. This is a remarkable result given the lack of freedom in the
spectral action to tune this term.

\end{abstract}
\maketitle










It has been known since the 1960's \cite{MTW} that in the Hamiltonian
quantization of gravity it is essential to include boundary terms in the
action, as this allows to define consistently the momentum conjugate to the
metric. This makes it necessary to modify the Einstein-Hilbert action by
adding to it a surface integral term so that the variation of the action
becomes well defined and yields the Einstein field equations. The reason for
this manipulation is that the curvature scalar $R$ contains second derivatives
of the metric, which are removed after integrating by parts to obtain an
action which is quadratic in first derivatives of the metric. These surface
terms are canceled by modifying the Euclidean action to \cite{Hawking},
\cite{HH}
\[
I=-\frac{1}{16\pi}%
{\displaystyle\int\limits_{M}}
d^{4}x\sqrt{g}R-\frac{1}{8\pi}%
{\displaystyle\int\limits_{\partial M}}
d^{3}x\sqrt{h}K,
\]
where $\partial M$ is the boundary of $M$, $h_{ab}$ is the induced metric on
$\partial M$ and $K$ is the trace of the second fundamental form on $\partial
M.$ We use the sign convention according to which $R$ is positive for the
sphere and $K$ is positive for the ball. Notice that there is a relative
factor of $2$ and a fixed sign between the two terms, and \ that the boundary
term has to be completely fixed. This is a delicate fine tuning and is not
determined by any symmetry, but only by the consistency requirement. There is
no known symmetry that predicts this combination and it is always added by hand.

In the noncommutative geometric approach to the formulation of a unified
theory of all fundamental interactions including gravity, the starting point
is the replacement of the Riemannian geometry of space-time with
noncommutative geometry. The basic data of noncommutative geometry consists of
an involutive algebra $\mathcal{A}$ of operators in Hilbert space
$\mathcal{H}$, which plays the role of the algebra of coordinates, and a
self-adjoint operator $D$ in $\mathcal{H}$ \cite{Connes} which plays the role
of the inverse of the line element. The spectrum of the standard model
indicates that the algebra is to be taken as $\mathcal{A}=C^{\infty}\left(
M\right)  \otimes\mathcal{A}_{F}$ where the algebra $\mathcal{A}_{F}$ is
finite dimensional, $\mathcal{A}_{F}=\mathbb{C}\oplus\mathbb{H}\oplus
M_{3}\left(  \mathbb{C}\right)  ,$ and $\mathbb{H}\subset M_{2}\left(
\mathbb{C}\right)  $ is the algebra of quaternions. The algebra $\mathcal{A}$
is a tensor product which geometrically corresponds to a product space. The
spectral geometry of $\mathcal{A}$ is given by the product rule
\[
\mathcal{H}=L^{2}\left(  M,S\right)  \otimes\mathcal{H}_{F},\quad
D=D_{M}\otimes1+\gamma_{5}\otimes D_{F},
\]
where $L^{2}\left(  M,S\right)  $ is the Hilbert space of $L^{2}$ spinors, and
$D_{M}$ is the Dirac operator of the Levi-Civita spin connection on $M.$ The
Hilbert space of quarks and leptons fixes the choice of the Dirac operator
$D_{F}$ and the action of $\mathcal{A}_{F}$ in $\mathcal{H}_{F}.$ The operator
$D_{F}$ anticommutes with the chirality operator $\gamma_{F}$ on
$\mathcal{H}_{F}.$ The spectral geometry does not change if one replaces $D$
by the equivalent operator
\begin{equation}
D=D_{M}\otimes\gamma_{F}+1\otimes D_{F}, \label{prod}%
\end{equation}
but this equivalence fails when $M$ has a boundary and it is only the latter
choice which has conceptual meaning since $\gamma_{5}$ no longer anticommutes
with $D_{M}$ when $\partial M\neq\emptyset$. The noncommutative space defined
by a spectral triple has to satisfy the basic axioms of noncommutative
geometry. This approach shares a common feature with Euclidean quantum gravity
in that the Riemannian manifold is taken to be Euclidean in order for the line
element, which is the inverse of the Dirac operator, to be compact. It is then
assumed that one obtains the Lorentzian results by analytically continuing the
expressions obtained by performing the path integral to Minkowski space. A
fundamental principle in the noncommutative approach is that the usual
emphasis on the points $x\in M$ of a geometric space is now replaced by the
spectrum of the operator $D.$ The spectral action principle states that the
physical action depends only the spectrum of the Dirac operator, which is
geometrical. Indeed, it was shown that all the fundamental interactions
including gravity are unified in the spectral action \cite{ACAC}
\[
I=\text{Tr}f\left(  \frac{D}{\Lambda}\right)  +\left\langle \Psi
,D\Psi\right\rangle ,
\]
where Tr is the usual trace of operators in the Hilbert space $\mathcal{H}$,
$\Lambda$ is a cut-off scale and $f$ \ is a positive function. The action is
then uniquely defined and the only arbitrariness one encounters is in the
first few coefficients in the spectral expansion since higher coefficients are
suppressed by the high-energy scale. This remarkable action includes the
gravitational Einstein-Hilbert term with the square of the Weyl tensor, the
$SU(3)_{c}\times SU(2)_{w}\times U(1)_{Y}$ gauge interactions, the Higgs
couplings including the spontaneous symmetry breaking, all coming with the
correct signs as well as a relation between the gauge couplings and Higgs
couplings. The geometrical model is valid at the unification scale, and
relates the gauge coupling constants to each other and to the Higgs coupling.
When these relations are taken as boundary conditions valid at the unification
scale in the renormalization group (RG) equations, one gets a prediction of
the Higgs mass to be around $170\pm10$ GeV, the error being due to our
ignorance of the physics at unification scale. In addition there is one
relation between the sum of the square of fermion masses and the $W$ particle
mass square which enables us to predict the top quark mass compatible with the
measured experimental value. It also accommodates small neutrino masses
through the see-saw mechanism, thanks to a more subtle choice (\cite{ACM}) of
the chirality operator $\gamma_{F}$ which gives to the geometry $F$ a
$KO$-dimension which is congruent to $6$ modulo $8$. The charge conjugation
operator $J$ \ for the product geometry \eqref{prod} is then given by
\[
J=J_{M}\,\,\gamma_{5}\otimes J_{F}%
\]
which commutes with the operator $D$ given by \eqref{prod} since in even
dimension $J_{M}$ commutes with $D_{M}$ while in dimension $6$ modulo $8$,
$J_{F}$ anticommutes with $\gamma_{F}$.

The results were derived for manifolds without boundary. We stress that
definition of the noncommutative space corresponding to the physical
space-time must satisfy the restrictive axioms of noncommutative geometry.
Once this is done, there is essentially no freedom left in determining the
spectral action, except for the three coefficients of the Mellin transform of
the function $f.$ These correspond to the cosmological constant, the Newton
constant and the gauge couplings and where the dependence on the energy scale
is governed by the renormalization group equations. Because of these
constraints, it is essential to find out whether the boundary terms of the
spectral action agree with the modifications dictated by the consistency of
quantum gravity. This is a severe test of the spectral action principle as
there is no freedom present in tuning the surface terms to reproduce the
desired results with correct signs and numerical values. It is the purpose of
this work to show that the spectral action does pass all tests predicting the
correct modification of the boundary terms. We can go further and make the
mass scale $\Lambda$ appearing in the Dirac operator dynamical by replacing it
with a dilaton field. \ We have recently shown that in this case the spectral
action becomes almost scale invariant and gives the same low-energy limit as
the Randall-Sundrum model as well as providing a model for extended inflation
\cite{Scale}. In other words, the simple form of the spectral action is
capable of producing all the desirable features of unified theories including
gravity with the correct physical predictions.

The Dirac operator in the spectral action must satisfy the hermiticity
condition
\[
\left\langle \Psi,D\Psi\right\rangle =\left\langle D\Psi,\Psi\right\rangle .
\]
These are satisfied provided the following "natural" boundary condition is
imposed \cite{Luck}, \cite{BG1},\cite{BG2}
\[
\Pi_{-}\Psi|_{\partial M}=0,
\]
where the projection operator $\Pi_{-}$ is given by $\Pi_{-}=\frac{1}%
{2}\left(  1-\chi\right)  $ where $\chi=\gamma_{n}\gamma_{5}$ satisfies
$\chi^{2}=1.$ The Clifford algebra is defined by $\left\{  \gamma^{\mu}%
,\gamma^{\nu}\right\}  =-2g^{\mu\nu}$ and we denote by $n$ the unit
\emph{inward} normal and $\gamma_{n}$ the corresponding Clifford
multiplication. Although one can keep the discussion general, it will be more
transparent to specialize to the case where the dimensions of the continuous
part of the noncommutative space is taken to be four. A local system of
coordinates on $M$ will be denoted by $x^{\mu}$, $\mu=1,\cdots4,$ and on
$\partial M$ will be denoted by $y^{a},$ $a=1,2,3.$ Let the functions $x^{\mu
}\left(  y^{a}\right)  $ be given by the embedding of the hypersurface in $M$
and let $e_{a}^{\mu}=\frac{\partial x^{\mu}}{\partial y^{a}},$ then the metric
$g_{\mu\nu}$ on $M$ induces a metric $h_{ab}$ on the hypersurface such that
$h_{ab}=g_{\mu\nu}e_{a}^{\mu}e_{b}^{\nu}$ and where $n^{\mu}$ is orthogonal to
$e_{a}^{\mu}$ so that $g_{\mu\nu}n^{\mu}e_{a}^{\nu}=0.$ It is convenient to
define $n_{\mu}=g_{\mu\nu}n^{\nu}$ so that $n_{\mu}e_{a}^{\mu}=0.$ We now
define the inverse functions $e_{\mu}^{a}$ by $e_{a}^{\mu}e_{\mu}^{b}%
=\delta_{a}^{b}$ which satisfies the condition $e_{a}^{\mu}e_{\nu}^{a}%
=\delta_{\nu}^{\mu}-n^{\mu}n_{\nu}$ to be consistent with $n_{\mu}e_{a}^{\mu
}=0.$ We therefore can write \cite{Poisson}
\[
g_{\mu\nu}=h_{ab}e_{\mu}^{a}e_{\nu}^{b}+n_{\mu}n_{\nu}.
\]
The inverse metric is also defined by $h^{ab}=g^{\mu\nu}e_{\mu}^{a}e_{\nu}%
^{b}$ and the inverse relation is
\[
g^{\mu\nu}=h^{ab}e_{a}^{\mu}e_{b}^{\nu}+n^{\mu}n^{\nu}.
\]
This shows that any tensor can be projected into the hypersurface using the
completeness relations for the basis $\left\{  e_{\mu}^{a},n_{\mu}\right\}
.$We finally define on $\partial M$, \
\[
\chi=-\frac{\sqrt{h}}{3!}\epsilon^{abc}\gamma_{a}\gamma_{b}\gamma_{c}%
,\quad\gamma_{5}=\chi\gamma_{n},
\]
which satisfy $\chi^{2}=1,$ $\chi\gamma^{a}=\gamma^{a}\chi,$ $\chi\gamma
^{n}=-\gamma^{n}\chi,$ $\gamma_{5}^{2}=1,$ $\chi\gamma_{5}=-\gamma_{5}\chi.$
The normal vector $n^{\mu}$ satisfies the properties%
\[
n_{\mu;\nu}=-K_{ab}e_{\mu}^{a}e_{\nu}^{b},\qquad e_{a;\nu}^{\mu}e_{b}^{\nu
}=\Gamma_{ab}^{c}e_{c}^{\mu}+K_{ab}n^{\mu}%
\]
where the covariant derivative $;\nu$ is the space-time covariant derivative
and $\Gamma_{ab}^{c}$ is the Christoffel connection of the metric $h_{ab}$,
and $K_{ab}$ is the extrinsic curvature whose symmetry follows from the
relation $e_{a;b}^{\mu}=e_{b;a}^{\mu}.$

The bosonic part of the spectral action is then obtained by using the identity
\cite{ACAC}%
\[
\text{Tr}\left(  f\left(  D^{2}/m^{2}\right)  \right)  \simeq%
{\displaystyle\sum\limits_{n\geq0}}
\,f_{4-n}\,a_{n}\left(  D^{2}/m^{2}\right)  ,
\]
where $f_{n}$ are related to the Mellin transforms of the function $f$. \ The
Seeley-deWitt coefficients $a_{n}\left(  P,\chi\right)  $ are geometrical
invariants. These were calculated for Laplacians which are the square of the
Dirac operator, for manifolds with boundary. To evaluate these terms, we first
write the Laplacian in the form
\begin{align*}
P  &  =D^{2}=-\left(  g^{\mu\nu}\partial_{\mu}\partial_{\nu}+\mathbb{A}^{\mu
}+\mathbb{B}\right) \\
&  =-\left(  g^{\mu\nu}\nabla_{\mu}^{^{\prime}}\nabla_{\nu}^{^{\prime}%
}+E\right)  ,
\end{align*}
where $\nabla_{\mu}^{^{\prime}}=\partial_{\mu}+\omega_{\mu}^{^{\prime}}$ and
$\omega_{\mu}^{^{\prime}}=\frac{1}{2}g_{\mu\nu}\left(  \mathbb{A}^{\nu
}+g^{\rho\sigma}\Gamma_{\rho\sigma}^{\nu}\right)  .$ It is convenient to write
the Dirac operator in the form
\[
D=\gamma^{\mu}\nabla_{\mu}-\Phi,
\]
where $\nabla_{\mu}=\partial_{\mu}+\omega_{\mu}$ and $\omega_{\mu}$ is the
torsion free spin-connection. The boundary conditions for $D^{2}$ are then
equivalent to \cite{BG1}, \cite{BG2}
\[
\mathcal{B}_{\chi}\Psi=\Pi_{-}\left(  \Psi\right)  |_{\partial M}\oplus\Pi
_{+}\left(  \nabla_{n}^{^{\prime}}+S\right)  \Pi_{+}\left(  \Psi\right)
|_{\partial M}=0,
\]
where
\begin{align*}
S  &  =\Pi_{+}\left(  \gamma_{n}\Phi-\frac{1}{2}\gamma_{n}\gamma^{a}\nabla
_{a}^{^{\prime}}\chi\right)  \Pi_{+},\\
\nabla_{a}^{^{\prime}}\chi &  =\partial_{a}\chi+\left[  \omega_{a}^{^{\prime}%
},\chi\right]  =K_{ab}\chi\gamma^{n}\gamma^{b}+\left[  \theta_{a},\chi\right]
,
\end{align*}
and where $\theta_{a}=\omega_{a}^{\prime}-\omega_{a}.$ We then have the
relations
\begin{align*}
E  &  =\gamma^{\mu}\nabla_{\mu}\Phi-\Phi^{2}-\frac{1}{2}\gamma^{\mu\nu}%
\Omega_{\mu\nu},\\
\Omega_{\mu\nu}  &  =\partial_{\mu}\omega_{\nu}^{^{\prime}}-\partial_{\nu
}\omega_{\mu}^{^{\prime}}+\omega_{\mu}^{^{\prime}}\omega_{\nu}^{^{\prime}%
}-\omega_{\nu}^{^{\prime}}\omega_{\mu}^{^{\prime}}.
\end{align*}
We list the first relevant Seeley-deWitt coefficients for Laplacians which are
square of Dirac operators \cite{Vass}
\[
a_{0}\left(  P,\chi\right)  =\frac{1}{16\pi^{2}}%
{\displaystyle\int\limits_{M}}
d^{4}x\sqrt{g}\text{Tr}\left(  1\right)  ,
\]%
\[
a_{1}\left(  P,\chi\right)  =0,
\]%
\begin{align*}
a_{2}\left(  P,\chi\right)   &  =\frac{1}{96\pi^{2}}\left(
{\displaystyle\int\limits_{M}}
d^{4}x\sqrt{g}\text{Tr}\left(  6E+R\right)  \right. \\
&  \qquad\left.  +%
{\displaystyle\int\limits_{\partial M}}
d^{3}x\sqrt{h}\text{Tr}\left(  2K+12S\right)  \right)  ,
\end{align*}%
\begin{align*}
a_{3}\left(  P,\chi\right)   &  =\frac{1}{384(4\pi)^{\frac{3}{2}}}%
{\displaystyle\int\limits_{\partial M}}
d^{3}x\sqrt{h}\text{Tr}\left(  96\chi E+3K^{2}\right. \\
&  \left.  +6K_{ab}K^{ab}+96SK+192S^{2}-12\nabla_{a}^{^{\prime}}\chi
\nabla^{^{^{\prime}a}}\chi\right)  ,
\end{align*}
As a warm up, these results could be applied to the simple case of an ordinary
Dirac operator
\[
D=\gamma^{\mu}\left(  \partial_{\mu}+\omega_{\mu}\right)  .
\]
Therefore, in the above formulas we have
\begin{align*}
\omega_{\mu}^{^{\prime}}  &  =\omega_{\mu},\quad E=-\frac{1}{4}R,\quad
\Phi=0,\\
S  &  =-\frac{1}{2}K\Pi_{+},\quad\nabla_{a}^{^{\prime}}\chi=K_{ab}\chi
\gamma^{n}\gamma^{b}%
\end{align*}
Substituting Tr$\left(  1\right)  =4$ and Tr$\left(  S\right)  =-K$ we have
for the first few terms
\[
a_{0}\left(  P,\chi\right)  =\frac{1}{4\pi^{2}}%
{\displaystyle\int\limits_{M}}
d^{4}x\sqrt{g}%
\]%
\[
a_{2}\left(  P,\chi\right)  =-\frac{1}{24\pi^{2}}\left(
{\displaystyle\int\limits_{M}}
d^{4}x\frac{1}{2}\sqrt{g}R+%
{\displaystyle\int\limits_{\partial M}}
d^{3}x\sqrt{h}K\right)
\]%
\[
a_{3}\left(  P,\chi\right)  =\frac{1}{32(4\pi)^{\frac{3}{2}}}%
{\displaystyle\int\limits_{\partial M}}
d^{3}x\sqrt{h}\left(  K^{2}-2K_{ab}K^{ab}\right)
\]

The important point in the above result is the emergence of the combination
\cite{Hawking}
\[
-%
{\displaystyle\int\limits_{M}}
d^{4}x\sqrt{g}R-2%
{\displaystyle\int\limits_{\partial M}}
d^{3}x\sqrt{h}K
\]
as the lowest term of the gravitational action which is known to be the
required correction to the Einstein action involving the surface term so as to
make the Hamiltonian formalism consistent. This is remarkable because both the
sign and the coefficient are correct. The only assumption made is that normal
boundary conditions are taken such that they enforce the hermiticity of the
Dirac operator. This is yet another miracle concerning correct signs obtained
in the spectral action of the Dirac operator. We also notice that the relative
coefficient between $R$ and $K$ depends, in general, on the nature of the
Laplacian. The desired answer is true for the square of the Dirac operator,
but \textit{not} for a general Laplacian. We note that there other boundary
conditons may lead to different results \cite{Vass}.

This is a general result and applies to all noncommutative models based on
spaces which are the tensor product of the spectral triple of a Riemannian
manifold by that of a discrete space. In particular the above feature also
works for the spectral action of the standard model. Indeed by applying the
above formulas to the Dirac operators in the quarks and leptonic sectors with
the corresponding boundary conditions one derives the full spectral action
with boundary terms included. We just give the results here; the full details
will appear in the expanded version of this letter \cite{prepare}. (Note that
in \cite{ACM} we use the opposite sign convention for the scalar $R$ ):
\begin{align*}
&  I=\frac{48\Lambda^{4}}{\pi^{2}}f_{4}%
{\displaystyle\int\limits_{M}}
d^{4}x\sqrt{g}\\
&  +\frac{8\Lambda^{2}}{\pi^{2}}f_{2}\left\{
{\displaystyle\int\limits_{M}}
d^{4}x\sqrt{g}\left(  -\frac{1}{2}R-\frac{1}{4}\left(  a\left\vert
\varphi\right\vert ^{2}+\frac{1}{2}c\right)  \right)  \right.  \\
&  \qquad\qquad\left.  -%
{\displaystyle\int\limits_{\partial M}}
d^{3}x\sqrt{h}K\right\}  \\
&  +\frac{2\Lambda}{(4\pi)^{\frac{3}{2}}}f_{1}%
{\displaystyle\int\limits_{\partial M}}
d^{3}x\sqrt{h}\left(  3\left(  K^{2}-2K_{ab}K^{ab}\right)  \right)
\end{align*}%
\begin{align*}
&  +\frac{f_{0}}{2\pi^{2}}\left\{
{\displaystyle\int\limits_{M}}
d^{4}x\sqrt{g}\left(  -\frac{3}{5}C_{\mu\nu\rho\sigma}^{2}+\frac{11}%
{30}R^{\ast}R^{\ast}-\frac{2}{5}R_{;\mu}^{\mu}\right.  \right.  \\
&  \qquad\qquad+a\left\vert D_{\mu}\varphi\right\vert ^{2}+\frac{1}{6}R\left(
a\left\vert \varphi\right\vert ^{2}+\frac{1}{2}c\right)  \\
\qquad &  \qquad\qquad\left.  +g_{3}^{2}\left(  G_{\mu\nu}^{i}\right)
^{2}+g_{2}^{2}\left(  F_{\mu\nu}^{\alpha}\right)  ^{2}+\frac{5}{3}g_{1}%
^{2}\left(  B_{\mu\nu}\right)  ^{2}\right)  \\
&  \qquad\qquad\left.  +b\left\vert \varphi\right\vert ^{4}+2e\left\vert
\varphi\right\vert ^{2}+\frac{1}{2}d-\frac{1}{3}a\left(  \left\vert
\varphi\right\vert ^{2}\right)  _{;\mu}^{\mu}\right\}  \\
&  +\frac{f_{0}}{2\pi^{2}}\left\{
{\displaystyle\int\limits_{\partial M}}
d^{3}x\sqrt{h}\left(  \frac{1}{3}K\left(  a\left\vert \varphi\right\vert
^{2}+\frac{1}{2}c\right)  \right.  \right.  \\
&  \left.  +\frac{2}{15}\left(  5RK+4KR_{\;nan}^{a}+4K_{ab}R_{\;acb}%
^{c}+18R_{anbn}K^{ab}\right)  \right)  \\
&  \qquad+\frac{4}{315}\left(  17K^{3}+39KK_{ab}K^{ab}-116K_{a}^{\;b}%
K_{b}^{\;c}K_{c}^{\;a}\right)  \},
\end{align*}
where $f_{n}=%
{\displaystyle\int\limits_{0}^{\infty}}
v^{n-1}f(v)dv,$ and
\[%
\begin{array}
[c]{c}%
a=\,\mathrm{tr}\left(  3\left\vert k^{u}\right\vert ^{2}+3\left\vert
k^{d}\right\vert ^{2}+\left\vert k^{e}\right\vert ^{2}+\left\vert k^{\nu
}\right\vert ^{2}\right)  ,\\
b=\,\mathrm{tr}\left(  3\left\vert k^{u}\right\vert ^{4}+3\left\vert
k^{d}\right\vert ^{4}+\left\vert k^{e}\right\vert ^{4}+\left\vert k^{\nu
}\right\vert ^{4}\right)  ,\,\\
c=\,\mathrm{tr}\left(  \left\vert k^{\nu_{R}}\right\vert ^{2}\right)  ,\qquad
d=\mathrm{tr}\left(  \left\vert k^{\nu_{R}}\right\vert ^{4}\right)  ,\\
e=\mathrm{tr}\left(  \left\vert k^{\nu_{R}}\right\vert ^{2}\left\vert k^{\nu
}\right\vert ^{2}\right)
\end{array}
\]
In the above expression, $g_{1}$, $g_{2}$, and $g_{3}$ are the $U(1),$ $SU(2)$
and $SU(3)$ gauge couplings with the corresponding gauge field strengths
$B_{\mu\nu},$ $F_{\mu\nu}^{\alpha}$ and $G_{\mu\nu}^{i}$, \ and where the
Higgs doublet is $\varphi$ and the Yukawa fermionic couplings are given by the
$3\times3$ matrices $k^{u},$ $k^{d},k^{e},$ $k^{\nu}$ and $k^{\nu_{R}}.$ The
first few boundary terms depend only on the gravitational fields, while the
Higgs field would begin to appear in the $a_{4}$ term. Contributions of the
vector fields drop out completely if we make the assumption that their normal
components vanish on the boundary: $A_{n}|_{\partial M}=0.$ Remarkably the
terms $\frac{1}{6}R\left(  a\left\vert \varphi\right\vert ^{2}+\frac{1}%
{2}c\right)  $ and $\frac{1}{3}K\left(  a\left\vert \varphi\right\vert
^{2}+\frac{1}{2}c\right)  $ appear again with the same sign and the same
relative factor of 2. This is a proof that the spectral action takes care of
its self consistency.

From all these considerations we deduce that the simple requirement of having
boundary conditions consistent with the hermiticity of the Dirac operator, is
enough to guarantee that the spectral action has all the correct features and
expected terms, including correct signs and coefficients.

Finally we note that we can include the effects of introducing a dilaton field
to make the mass scale dynamical and obtain an almost scale invariant action.
The main results obtained recently \cite{Scale} where it was shown that the
dilaton interacts only through its kinetic term with a potential generated at
the quantum level. The model has the same low-energy sector as the
Randall-Sundrum model and the model of extended inflation. In the case of
manifolds without boundary, the only modifications needed in the spectral
action is the addition of the dilaton terms $\frac{8}{3\pi^{2}}f_{2}\int
_{M}d^{4}x\sqrt{G}G^{\mu\nu}\partial_{\mu}\phi\partial_{\nu}\phi$. For
manifolds with boundary there will be additional terms and these could play
some role in cosmological considerations.

\begin{acknowledgments}
The research of A. H. C. is supported in part by the National Science
Foundation under Grant No. Phys-0601213, and by the Arab Fund for Economic and
Social Development.
\end{acknowledgments}

\end{document}